\documentclass[twocolumn,aip,prb,showpacs,floatfix]{revtex4}

\usepackage{epsf}
\usepackage[dvips]{graphicx}
\usepackage{bm}
\begin{document}
    
\title{
Geometric, electronic properties and the thermodynamics\\
of pure and Al--doped Li clusters
}

\author {Mal--Soon Lee$^1$}  
\author {S. Gowtham$^2$}    
\author {Haiying He$^2$}   
\author {Kah--Chun Lau$^2$} 
\author {Lin Pan$^2$}       
\author {D. G. Kanhere$^1$} 

\affiliation{
$^1$ Centre for Modeling and Simulation, Department of Physics, 
University of Pune, Ganeshkhind, Pune 411 007, India.\\ 
$^2$ Physics Department, Michigan Technological University, 
Houghton, MI 49931-1295, USA.}

\date{\today}

\begin{abstract} 
The first--principles density functional molecular dynamics simulations have been carried out
to investigate the geometric, the electronic, and the finite temperature properties of pure 
Li clusters (Li$_{10}$, Li$_{12}$) and Al--doped Li clusters (Li$_{10}$Al, Li$_{10}$Al$_2$).
We find that addition of two Al impurities in Li$_{10}$ results in a substantial
structural change, while the addition of one Al impurity causes a rearrangement of atoms.
Introduction of Al--impurities in Li$_{10}$ establishes a polar bond between 
Li and nearby Al atom(s), leading to a multicentered bonding, which weakens 
the Li--Li metallic bonds in the system.
These weakened Li--Li bonds lead to a premelting feature
to occur at lower temperatures in Al--doped clusters. 
In Li$_{10}$Al$_2$, Al atoms also form a weak covalent bond, resulting into their dimer like 
behavior.
This causes Al atoms not to `melt' till 800~K, in contrast to the Li atoms which show 
a complete diffusive behavior above 400~K.
Thus, although one Al impurity in Li$_{10}$ cluster does not change its melting 
characteristics significantly, two impurities results in `surface
melting' of Li atoms whose motions are confined around Al dimer. 

\end{abstract} 
\pacs{61.46.Bc, 36.40.--c, 36.40.Cg, 36.40.Ei}  

\maketitle

\section{Introduction \label{intro}}

There have been considerable experimental and theoretical studies 
to understand the physical and chemical properties of cluster 
which include structural and electronic properties, the nature of bonding, 
thermodynamics,~\cite{Francesca} and spectroscopic properties.
Since most of these studies have been carried out on the homogeneous 
clusters,~\cite{homo} the physics of mixed clusters remains less explored.~\cite{hetero}
It is well known that the dilute impurities alter the electronic structure 
and geometries of the bulk system.
A similar phenomenon is also observed in impurity--doped clusters.~\cite{LiSn, Li6Sn}

Cheng {\it et al.} have investigated the energetics and the electronic structure of
small Li-Al clusters.~\cite{Landman}
They suggested a special role of AlLi$_5$ unit as a building block 
for clusters of assembled materials, {\it e.g.} Al$_n$Li$_{5n}$ ($n>1$).
This idea was further explored by Akola and Manninen.~\cite{Akola}
They investigated small Li--rich Al$_N$Li$_{5N}$ ($N$=1-6,10) clusters using
first--principle calculations.
They reported that Al ions form a compact inner core embedded in Li atoms. 
However, they did not find AlLi$_5$ to be a favorable
candidate as a building block for larger clusters.
Further, they observed a significant charge transfer from Li to
nearby Al atoms, strengthening ionic bonds between Li and Al, as well as a
formation of Al--Al covalent bonds.
Similar findings on Li$_{10}$Al$_8$ cluster have been reported by Kumar.~\cite{Kumar}

Another issue of considerable interest is the finite temperature behavior of homogeneous
as well as impurity--doped clusters. 
Joshi {\it et al.} have investigated the finite temperature
behavior of impurity--doped cluster, Li$_6$Sn.~\cite{Li6Sn}
Their work indicates that the addition of one impurity results in lowering of
melting temperature by about 125~K.
A similar observation has made be Aguado {\it et al.} for the case of LiNa$_{54}$ and 
CsNa$_{54}$ clusters.~\cite{Aguado-impurity}
Recently, an interesting study on the effect of a single impurity in the 
icosahedral clusters of silver has been reported by Mottet {\it et al.}~\cite{Mottet} 
In contrast to the previous studies,~\cite{Li6Sn, Aguado-impurity}
they showed that a single impurity of Ni or Cu can lead to an
increase in the melting temperature of the host.
A recent study by Zorriassatein {\it et al.} on the melting of Si$_{16}$Ti also 
reveals that a single impurity like Ti can change the finite temperature behavior of the 
host cluster significantly.~\cite{Si16Ti}
Therefore, it is of considerable interest to investigate the impurity induced
effects in Li clusters.
In the present work, we have carried out the first--principles density functional  
molecular dynamics simulations on Li$_{10}$Al and Li$_{10}$Al$_2$ clusters.
The results have been compared with those of pure Li clusters.
In particular, we have investigated the equilibrium geometries, the nature of bonding, and 
the finite temperature behavior of these clusters.
In sec. II, we briefly describe the computational procedure.
We present our results and discussion in sec. III. 
A brief summary of results is given in sec. IV. 

\section{Computational Details \label{comp}}

We have employed Bohn--Oppenheimer molecular dynamics~\cite{BOMD}
using Vanderbilt's ultrasoft pseudopotentials~\cite{Pseudo} within the local 
density approximation, as implemented in the VASP package.~\cite{VASP}
A cubic supercell of length 20~\AA\ with energy cutoff of 9.5~Ry was
used for the total--energy convergence.  
We have obtained the lowest energy structure and other equilibrium geometries by a two step 
process.
In the first step, starting from random configurations, the clusters are heated to 
a few representative temperatures, below and above expected melting points.
The clusters are maintained at these temperatures for at least 60~ps.
Then, resulting trajectories are used to choose several initial configurations for geometry 
optimizations.
In this way, we have obtained various equilibrium geometries.
The geometries are considered to be converged when the force
on each ion is less than 0.005~eV/\AA\ with a convergence
in the total energy to be the order of 10$^{-4}$~eV.
To investigate the nature of the bonding, we have examined the total charge
density and the molecular orbitals (MO's).

To examine the finite temperature behavior of clusters, 
molecular dynamics simulations have been carried out at 16 temperatures 
for Li$_{10}$, and at 13 temperatures for Li$_{12}$ and Li$_{10}$Al$_m$ ($m$=1,2) 
within the range of 100~K $\le$ T $\le$ 800~K.
The simulation time for each temperature is at least 150~ps.
We have discarded the first 30~ps for each temperature to allow the system 
to be thermalized.
The resulting trajectories have been used to calculate standard thermodynamic 
indicators as well as the ionic specific heat via multiple histogram technique.
The details can be found in Ref.~\onlinecite{SiSn}.

\section{Results and Discussion \label{results}}

\subsection{\bf {The Geometries and the Electronic Structures}}

\begin{figure}
  \epsfxsize 2.8in
  \epsffile{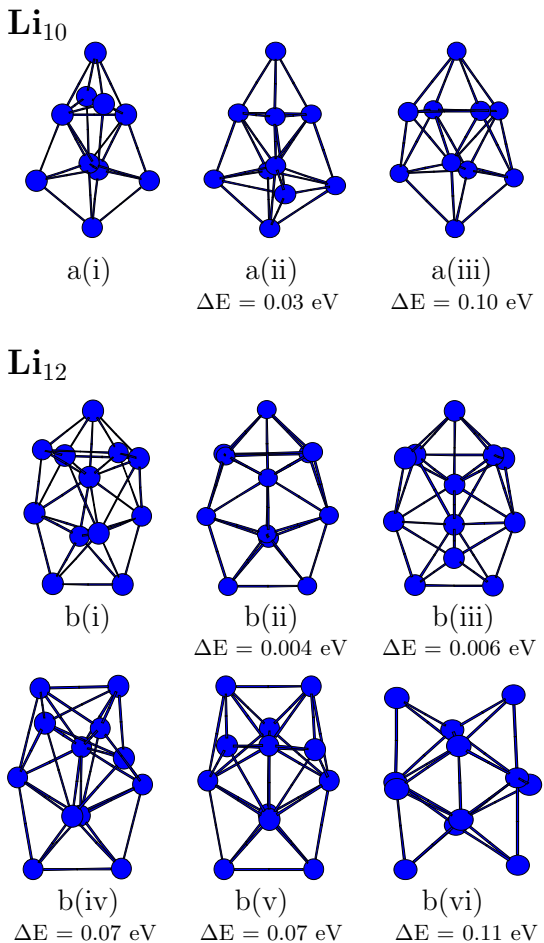}
  \caption{The ground--state geometry and some isomers of Li$_{10}$ and Li$_{12}$.
   The label (i) represents the ground--state geometry.  
   The energy difference $\Delta E$ is given in eV with respect to the 
   ground--state energy.}
  \label{fig1-1}
\end{figure}
\begin{figure}
  \epsfxsize 3.2in
  \epsffile{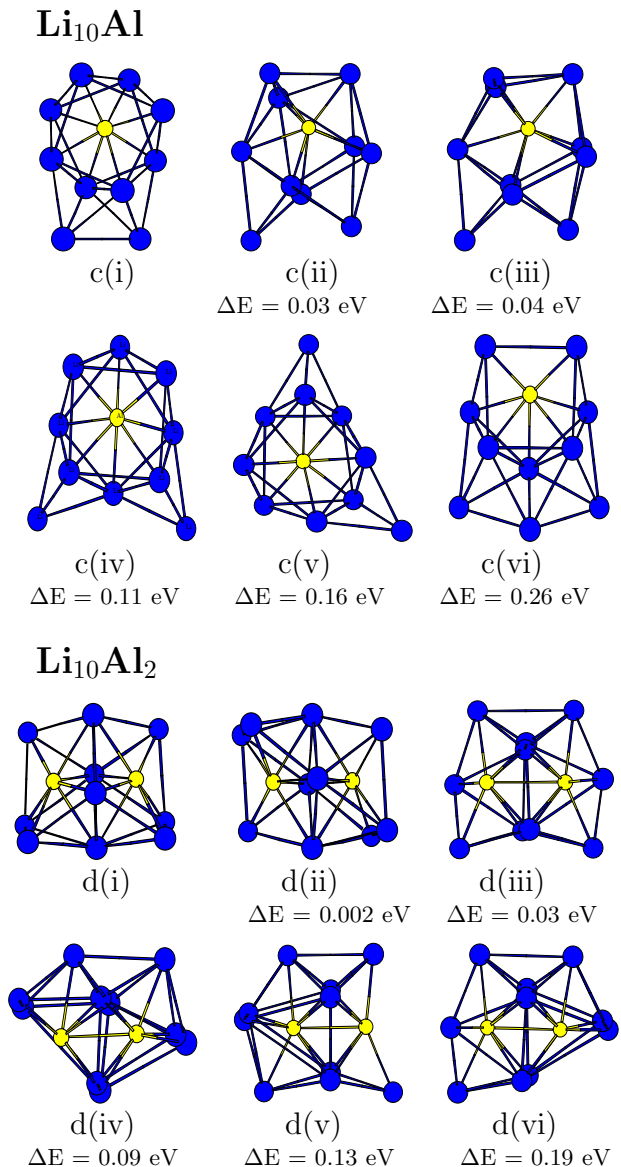}
  \caption{The ground--state geometry and some isomers of 
   Li$_{10}$Al and Li$_{10}$Al$_2$.
   The label (i) represents the ground--state geometry.  
   The blue circle represents the Li atoms and the yellow circle represents 
   the Al atoms.
   The energy difference $\Delta E$ is given in eV with respect to the 
   ground--state energy.}
  \label{fig1-2}
\end{figure}
The lowest energy structures and various isomers of Li$_{10}$, Li$_{12}$, Li$_{10}$Al, 
and Li$_{10}$Al$_2$ are shown in Figs.\ \ref{fig1-1} and \ref{fig1-2}.
The ground state (GS) geometry of Li$_{10}$ consists of two 
interconnected pentagonal rings with planes perpendicular to each other 
(Fig.\ \ref{fig1-1}--a(i)), which agrees with that found by 
Fournier {\it et al.}~\cite{Li-Fournier} as well as with our earlier study.~\cite{LiSn}
The structure with an atom at the center (Fig.\ \ref{fig1-1}--a(ii)) is nearly
degenerate with the ground state (Fig.\ \ref{fig1-1}--a(i)).
This is the structure reported by Jones {\it et al.} as their ground state.~\cite{Li10}
A high energy structure shows an interconnection of dodecahedron and 
octahedron (Fig.\ \ref{fig1-1}--a(iii)). 
We have found five nearly degenerate equilibrium geometries of Li$_{12}$ within the energy 
range of 0.007~eV.
We examined the stability of these structures by vibrational analysis. 
Three of these are shown in Figs.\ \ref{fig1-1}--b(i) $\sim$ \ref{fig1-1}--b(iii),
where Fig.\ \ref{fig1-1}--b(ii) have been reported as the lowest--energy structure by 
Fournier {\it et al.}~\cite{Li-Fournier} 
All the structures consist of two units: a pentagonal bipyramid and a distorted octahedron
connecting to each other with different angles.
As we shall see, the existence of these nearly degenerate structures
have a bearing on the shape of the specific heat curve at low temperatures.
In high energy structures, the octahedron is destroyed first 
(Figs.\ \ref{fig1-1}--b(iv) and \ref{fig1-1}--b(v)) and then the destruction of 
pentagonal bipyramid follows (Fig.\ \ref{fig1-1}--b(vi)).

Now, we discuss the geometries of impurity--doped systems.
So far as Li$_{10}$ is concerned, a single Al atom replaces one of the Li atoms in one 
unit of the pentagonal bipyramids in the GS geometry of Li$_{10}$Al (Fig.\ \ref{fig1-2}--c(i)).
This causes the Li atoms in another unit to rearrange in the form of antiprism, with Al atom 
at the center.
As shown in Figs.\ \ref{fig1-2}--c(ii) to \ref{fig1-2}--c(v), the  
low energy geometries are dominated by the presence of pentagonal bipyramid, 
whereas the high energy structures are dominated by antiprism.
The addition of one more Al atom in Li$_{10}$Al changes the lowest energy structure
significantly.
The equilibrium structures of Li$_{10}$Al$_2$ do not show the pentagonal bipyramidal 
structure any more.
Instead, there is an octahedron consisting of four Li atoms and two Al atoms with Li atoms  
forming a central plane.
This core octahedron is common to all the isomers.
Different isomers represent different ways of capping this core by the remaining Li atoms.
Symmetric arrangements of Li atoms give rise to low energy isomers (Fig.\ 
\ref{fig1-2}--d(i) $\sim$ \ref{fig1-2}--d(iii)).
The lowest energy structure (Fig.\ \ref{fig1-2}--d(i)) is the most symmetric structure 
among all the clusters studied.
This is because two Al atoms share Li atoms equally to fill their
unoccupied {\it p}--orbitals.
The nearly degenerate structure shown in Fig.\ \ref{fig1-2}--d(ii) has been found 
as the lowest energy structure by Cheng {\it et al.}~\cite{Landman} 
The stability of two nearly degenerate structures has been verified by carrying out 
a vibrational analysis.
The structures with broken symmetry have higher energies
(Figs.\ \ref{fig1-2}--d(iv) $\sim$ \ref{fig1-2}--d(vi)).

\begin{figure}
  \epsfxsize 2.8in
  \epsffile{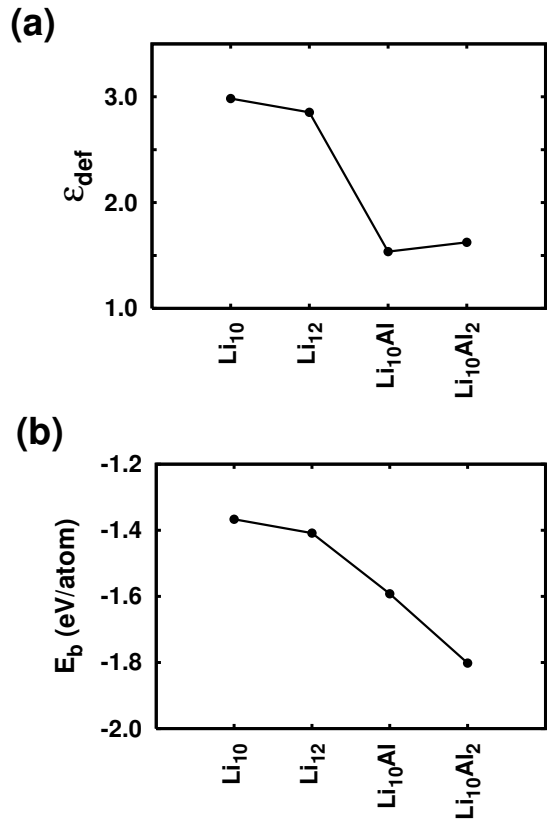}
  \caption{The properties of the ground state geometry. 
  (a) The deformation parameter ($\varepsilon_{def}$);
  (b) The binding energy E$_b$ in eV/atom.}
  \label{fig2}
\end{figure}
We used the deformation parameter $\varepsilon_{def}$ to examine the 
shape of the clusters.
$\varepsilon_{def}$ is defined as, 
$$
\varepsilon_{def} = {2Q_1 \over Q_2+Q_3}
$$
where $Q_1 \ge Q_2 \ge Q_3$ are eigenvalues of the quadrupole tensor
$$
Q_{ij} = \sum_I R_{Ii}R_{Ij}
$$
\noindent
with $R_{Ii}$ being {\it i}$^{th}$ coordinate of ion $I$ relative to the center of mass 
of the cluster.
A spherical system has $\varepsilon_{def}$ = 1 ($Q_1 = Q_2 = Q_3$), while 
$\varepsilon_{def} > $ 1 indicates a deformation.
The calculated $\varepsilon_{def}$ for the GS geometry of the clusters are 
shown in Fig.\ \ref{fig2}(a).
It can be seen that doping reduces the deformation considerably.
We also show the binding energies of these clusters in Fig.\ \ref{fig2}(b).
Clearly the increase in the binding energy by adding Al atom (0.2~eV/atom) 
is much higher than that by adding Li atom (0.04~eV/atom).
Thus, we conclude that the increase of the binding energy in the mixed clusters is mainly
due to the formation of strong Li--Al bonds.

\begin{figure}
  \epsfxsize 2.6in
  \epsffile{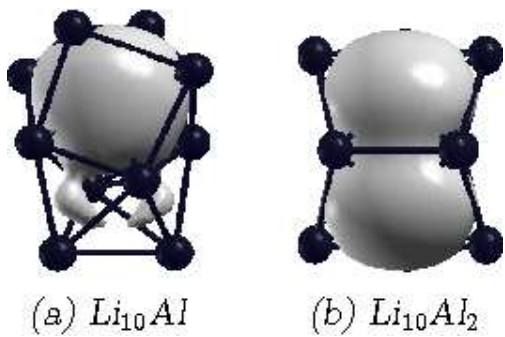}
  \caption{The isovalued surfaces of the total charge density at 1/3 of its 
   maximum values, where the maximum charge density of Li$_{10}$Al is 0.233 and that 
   of Li$_{10}$Al$_2$ is 0.239.}
  \label{fig3}
\end{figure}
\begin{figure}
  \epsfxsize 3.0in
  \epsffile{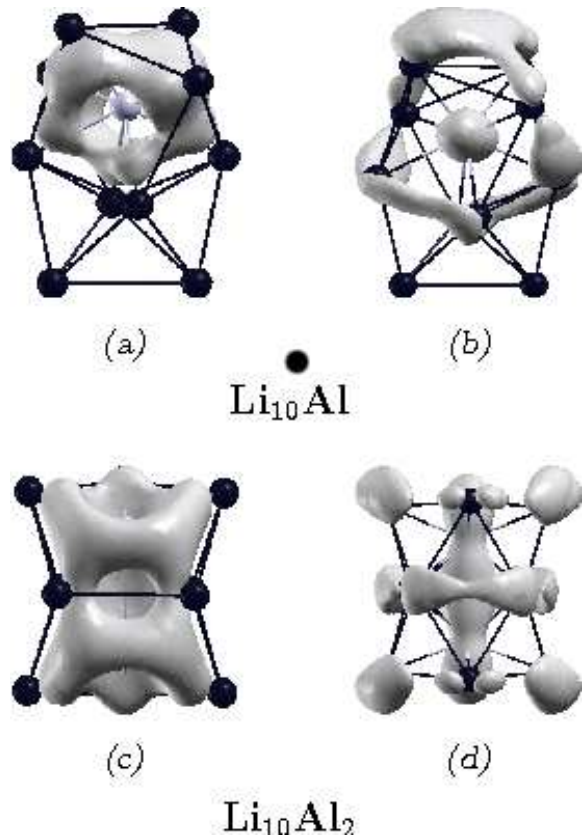}
  \caption{The difference charge density between the mixed cluster and separated units 
   of Li and Al clusters at the same positions. 
   (a) and (c) show the region where the charge is gained and 
   (b) and (d) show the region where the charge is lost.
   The figures are shown the isosurfaces of difference charge density at the value of 
   charge gain of 0.017 and loss of 0.008, respectively.}
  \label{fig4}
\end{figure}

We examine the change in the nature of the bonding due to the presence of Al atoms
via the total electron charge density, and the molecular orbitals (MO's).  
In addition, we have also calculated the difference charge density between electron 
charge densities of mixed cluster Li$_{10}$Al$_m$ ($m$=1,2) and separated units 
of Li$_{10}$ and Al$_m$ ($m$=1,2), by keeping the atomic positions the same as those in 
the mixed cluster. 
We first examine the charge distribution in Al$_2$ dimer and Li--Al diatomic cluster.
Evidently, there is a covalent bond in Al$_2$ dimer as expected, while Li--Al diatomic 
cluster shows a polar bond.
This can be explained on the basis of electronegativity difference between Li(0.98) 
and Al(1.61), which is not large enough to establish an ionic bond.
We note that the bond lengths of these clusters are 2.68~\AA\ for Al--Al and 
2.88~\AA\ for Li--Al.
The isovalued surface of total charge density in Li$_{10}$Al and Li$_{10}$Al$_2$ are 
shown in Figs.\ \ref{fig3}(a) and \ref{fig3}(b), respectively.
In Figs.\ \ref{fig4}(a) to \ref{fig4}(d), we show the isosurface of difference charge 
density for these Al--doped clusters.
When we add one Al atom to the Li$_{10}$ cluster, the total charge
distribution shows a mixed character of localization near Al atom and delocalization 
on the pentagonal bipyramid in Li$_{10}$Al (Fig.\ \ref{fig3}(a)).
It can be seen that the charge, localized around Al atom in the antiprism part, is 
nearly spherical. 
This is due to a significant charge transfer from nearby Li atoms to the Al atom 
to fill its unoccupied $p$--orbitals
However, there is an insignificant charge transfer from the Li atoms 
to the Al atom, forming the pentagonal bipyramidal unit. 
This leads to a delocalized charge distribution in this unit.
This observation is confirmed by the examination of Figs.\ \ref{fig4}(a) and 
\ref{fig4}(b), depicting the constant density contour of the difference charge density.
It may be noted that the values of the maxima in the difference charge density are 0.042 
and 0.051, and those of the minima are -0.037 and -0.084 in Li$_{10}$Al and 
Li$_{10}$Al$_2$, respectively.
Figs.\ \ref{fig4}(a) to \ref{fig4}(d) show the isosurfaces for the value of 0.017 for the charge 
gain, and -0.008 for the charge loss.
It can be seen that a ring shape region in between Li and Al (around Al) gains charge
in Li$_{10}$Al (Fig.\ \ref{fig4}(a)). 
Further, most of the charge is lost by Li atoms nearer to the Al atom (Fig.\ \ref{fig4}(b)).
This charge transfer from Li to Al causes Li atoms to be positively charged.
These positively charged Li atoms pull the charge distribution
around Al atom to polarize it.
This leads to a multicentered bonding between Al atom and nearby Li atoms. 
The addition of one more Al atom changes the charge density distribution from a mixed 
localized and delocalized one to a mainly localized one. 
In Li$_{10}$Al$_2$, the charge distribution (Fig.\ \ref{fig3}(b)) is mainly 
around two Al atoms polarized by nearby Li atoms.
The difference charge density shows a very symmetric charge gain region in between 
Li and Al atoms (Fig.\ \ref{fig4}(c)) and loss of charge at each atomic site 
(Fig.\ \ref{fig4}(d)).
These difference in the bonding between Li$_{10}$Al and 
Li$_{10}$Al$_2$ are also seen in molecular orbitals (figures not shown).
For example, in Li$_{10}$Al three highest occupied molecular orbital (HOMO) show 
a delocalized charge distribution, while only the HOMO shows a delocalization in Li$_{10}$Al$_2$.  
As noted earlier, since electronegativity difference of Li and Al is not very high, 
the Li atoms donate their charge to Al atom(s) partially.
We expect this charge transfer to result in weakening of Li--Li metallic bond 
in the system.
This feature has also been noted in Li$_6$Sn cluster.~\cite{Li6Sn}
Further, this partial charge transfer also results in the sharing of charge 
between two Al atoms to fill their unoccupied $p$--orbitals in Li$_{10}$Al$_2$.
As we shall see, this difference in the bonding leads to a different 
thermodynamic behavior in two clusters.

\begin{figure}
  \epsfxsize 2.8in
  \epsffile{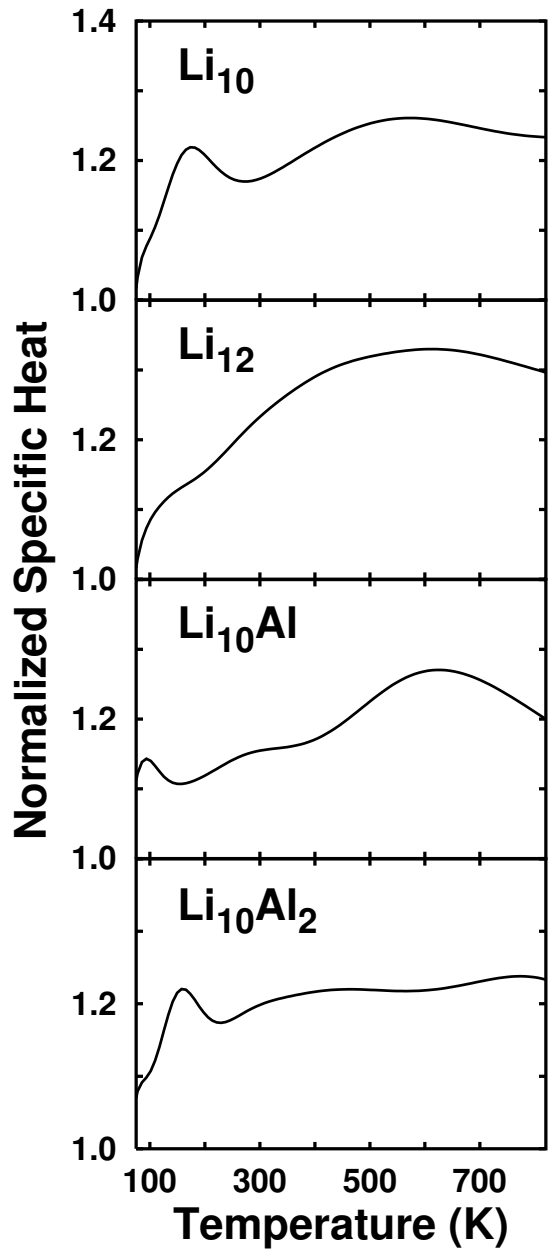}
  \caption{The normalized specific--heat as a function of temperature.
   $C_0=(3N-9/2)k_B$ is the zero--temperature classical limit of 
   the rotational plus vibrational canonical specific--heat.}
  \label{fig5}
\end{figure}
\subsection{Thermodynamics}
\subsubsection{Pure Li clusters}
\begin{figure}
  \epsfxsize 2.8in
  \epsffile{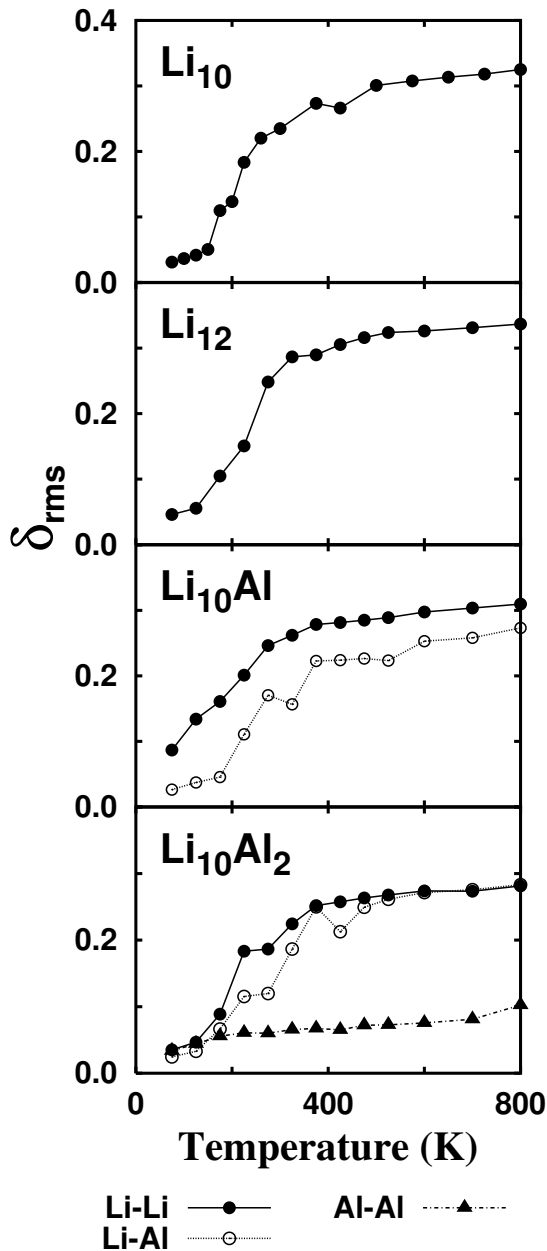}
  \caption{The root--mean--square bond length fluctuation ($\delta_{rms}$) of 
   Li--Li, Li--Al, Al--Al as a function of temperature.}
  \label{fig6}
\end{figure}

To investigate the melting behavior of the clusters, we calculate the canonical 
specific heat using multiple histogram technique.
The calculated ionic specific heats for all the clusters are shown in Fig.\ \ref{fig5}.
It is well known that a small cluster exibits a broad melting transition.
Our observatios are consistent with this observation.
In addition, in Li$_{10}$ the ionic specific heat also shows a remarkable premelting feature 
between 150~K and 225~K, whose maximum value is close to that of the main peak around 575~K.
The examination of the ionic motion reveals that the two structures shown in Figs.\
\ref{fig1-1}--a(i) and \ref{fig1-1}--a(ii) are observed in the temperature range of 150~K
$\sim$225~K.
The system visits these two isomers in the time spent of a few pico seconds at 175~K,
indicating that this premelting feature is due to the isomerization.
At higher temperature of 260~K, we observe the structure shown in Fig.\ \ref{fig1-1}--a(iii).
At still higher temperatures, the cluster visits these three isomers frequently.
It is difficult to identify the melting temperature for this cluster 
because liquidlike behavior develops over a wide range of temperature (300~K to 700~K). 
The root--mean--square bond length fluctuation ($\delta_{rms}$) is another indicator for 
studying a melting transition.
According to the Lindemann criteria (for bulk) solid--liquid transition is signified  
when the value of $\delta_{rms}$ exceeds 0.1.
However, it is generally observed that for clusters a liquidlike behavior is seen 
when $\delta_{rms}$ exceeds 0.25$\sim$0.3.
As shown in Fig.\ \ref{fig6}, $\delta_{rms}$ increases in two steps.
At 175~K it exceeds the value of 0.1, which corresponds to maximum specific heat of 
the shoulder.
However, as discussed earlier, the cluster is not in the liquidlike state.
$\delta_{rms}$ increases from 200~K again until it saturates about 0.3 at 575~K where we 
observe the maximum value of the specific heat.

The specific heat of Li$_{12}$ shows a weak shoulder around 125~K before it 
increases gradually.
This shoulder is related to the existence of five degenerate isomers.
At low temperatures up to 150~K, the motion is dominated by vibrations and the cluster 
visits the low energy isomers.
As temperature increases, the liquidlike behavior evolves over a very wide range of temperature.
At 175~K, we observe high energy isomers where the octahedral structure is absent (two of them 
are shown in Figs.\ \ref{fig1-1}--b(iv) and \ref{fig1-1}--b(v)).
With further increase of temperature to 375~K, we observe that the pentagonal bipyramidal 
structure is destroyed (one of these structures is shown in Fig.\ \ref{fig1-1}--b(vi)).
It can be seen that $\delta_{rms}$ also increases continuously from 125~K to 500~K 
(Fig.\ \ref{fig6}).
It is interesting to note that even though the isomerization is seen in both the clusters 
Li$_{10}$ and Li$_{12}$, evidently the isomerization in Li$_{10}$ leads to a prominent 
shoulder in its specific heat.
This is related to the nature of the process by which their ground state visits the isomer(s).
In the case of Li$_{10}$, there is a barrier at about 175~K.
The system overcomes this barrier with sudden increase in the accessible density of states.
For Li$_{12}$, since the different isomers differ only in the angle between two constituent units,
the ground state visits them almost continuously, aided by the vibrational motion.
This results in different specific heat curves. 

\subsubsection{Thermodynamics of Al-doped Li clusters}

In can be seen from Fig.\ \ref{fig5} that Li$_{10}$Al shows a premelting feature 
in the low temperature region around 150~K. 
This may be caused by a weakening of the Li--Li metallic bond due to the charge 
transfer from Li atoms to the Al atom in antiprism unit.
This effect has also been observed in Li$_6$Sn.~\cite{Li6Sn}
The Li atoms in this unit break the Li--Li bonds at low temperature 
and one of them moves towards the pentagonal bipyramidal unit.
This is the way the cluster visits the other low energy isomers shown in 
Figs.\ \ref{fig1-2}--c(ii) and \ref{fig1-2}--c(iii).
For the temperatures above 175~K, we observe the motion of Li atoms around the Al atom
without destroying the overall shape of the structure.
However, the four Li atoms consisting of pentagonal bipyramidal unit do not 
interchange their position with others till about 575~K.   
At this temperature, the system shows a diffusive liquidlike behavior.
We calculate the root--mean--square bond length fluctuation ($\delta_{rms}$) for 
Li--Li, Li--Al, and Al--Al separately, to see the difference between their fluctuations.
$\delta_{rms}$ is defined as, 
$$
\delta_{rms} = {1 \over N} \sum_{i<j}
               {\sqrt{\langle R_{ij}^{2} \rangle_{t}-
                      \langle R_{ij}     \rangle_{t}^{2}}
                  \over
                \langle R_{ij}\rangle_{t}}
$$
where $R_{ij}$ is the distance between $i^{th}$ and $j^{th}$ ions with $i$ and $j$ for 
relevant bonds, and $N$ is the number of bonds.
$\langle$...$\rangle_t$ denotes a time average over the entire trajectory.
It can be seen in Fig.\ \ref{fig6} that in the low temperature region $\delta_{rms}$ of Li--Al 
is much lower than that of Li--Li. 
This is because of relatively stronger Li--Al bond.

Interestingly, the specific heat of Li$_{10}$Al$_2$ at low temperature also shows 
a shoulder which is very similar to the specific heat of Li$_{10}$.
After this premelting feature, the specific heat remains flat.
This is due to the presence of Al dimer in the cluster.
When we compare its molecular orbitals with those of pure Al dimer, they looked alike 
even though their bond lengths are different.
The bond length of Al--Al in Li$_{10}$Al$_2$ is 2.97\AA, while that of 
pure Al dimer is 2.68\AA.
Thus, there is a weak covalent bond between Al atoms.
This covalent bond between them restricts the motion of Al atoms.
We observe that Al-Al bond does not break at least up to the temperature of about 800~K.
As shown in Fig.\ \ref{fig6}, $\delta_{rms}$ of Li--Li and Li--Al saturate 
to a value of about 2.5 at 500~K, while that of Al--Al reaches a value of 0.1 at 800~K.     
The saturation values of $\delta_{rms}$ for Li--Li and Li--Al are lower than 
corresponding $\delta_{rms}$ of other clusters, indicating the motion of atoms in 
Li$_{10}$Al$_2$ is more restricted than other systems at comparable temperature.
 
\section{Summary \label{summ}}
We have employed {\it ab initio} molecular dynamics to study the
equilibrium geometries, the electronic structure, and the finite temperature
properties of pure lithium clusters Li$_{10}$ and Li$_{12}$, and Al--doped
lithium clusters Li$_{10}$Al and Li$_{10}$Al$_2$.
We find that there is a substantial structural change upon doping with two Al atoms,
while the addition of one Al atom results in a rearrangement.
The analysis of the total charge density and the molecular orbitals
reveal that there is a partial charge transfer from Li atoms to Al atom(s) in Al--doped 
clusters, forming a polar bond between them.
This leads to a multicentered bonding and weakens the Li--Li metallic bonds in 
these clusters.
In Li$_{10}$Al$_2$, Al atoms also form a weak covalent bond.
These changes in the nature of bonding upon doping affect the finite temperature 
properties of pure host cluster.
We observe that the presence of dimer--like Al atoms with weak covalent
bond confines the motion of Li atoms around them.
Thus, a substitution of two Li atoms by Al atoms in Li$_{12}$ leads to a 
surface melting only, showing a continuous phase change over a very broad range of
temperature.

\section{Acknowledgments}
We acknowledge partial assistance from the Indo--French Center for
providing the computational support.
DGK acknowledges with pleasure, the hospitality of the Department of
Physics, MTU, USA during the course of this work.


\begin{thebibliography}{000}

\bibitem{Francesca}
   F. Baletto and R. Ferrando,
   Rev. Mod. Phys. {\bf 77}, 371 (2005).

\bibitem{homo}   
   H. W. Kroto, J. R. Heath, S. C.O'brian, 
   R. F. Curl, and R. E. Smalley, Nature, {\bf 318}, 162 (1985);  
   M. F. Jarrold and V. A. Constant, 
   Phys. Rev. Lett., {\bf 67}, 2994 (1991); 
   M. F. Jarrold and J. E. Bower, 
   J. Chem. Phys., {\bf 96}, 9180 (1992); 
   J. M. Hunter, J. L. Fye, M. F. Jarrold, and
   J. E. Bower, Phys. Rev. Lett., {\bf 73}, 2063 (1994); 
   A. A. Shvartsburg and M. F. Jarrold, 
   Phys. Rev. A, {\bf 60}, 1235 (1999);
   C. Majumder, V. Kumar, H. Mizuseki, and Y. Kawazoe, 
   Phys. Rev. B, {\bf 64}, 233405 (2001).

\bibitem{hetero}   
   R. O. Jones, A. I. Lichtenstein, and J. Hutter,
   J. Chem. Phys., {\bf 106}, 4566 (1997); 
   M. Deshpande, A. Dhavale, R. R. Zope, S. Chacko, and 
   D. G. Kanhere, Phys. Rev. A, {\bf 62}, 063202 (2000);
   Rajendra R. Zope, S. A. Blundell, Tunna Baruah, and 
   D. G. Kanhere, J. Chem. Phys., {\bf 115}, 2109 (2001);
   M. Deshpande, D. G. Kanhere, Igor Vasiliev and 
   R. M. Martin, Phys. Rev. A, {\bf 65}, 033202 (2002);
   S. Shetty, D. G. Kanhere, and S. Pal,
   J. Phys. Chem. A {\bf 108}, 628 (2004); 
   S. Chacko, D. G. Kanhere, and V. V. Paranjape,
   Phys. Rev. A {\bf 70}, 023204 (2004).

\bibitem{LiSn}
   M.--S. Lee, D. G. Kanhere, and Kavita Joshi,
   Phys. Rev. A {\bf 72}, 015201 (2005).

\bibitem{Li6Sn}
   K. Joshi and D. G. Kanhere,
   J. Chem. Phys. {\bf 119}, 12301 (2003).

\bibitem{Landman}
   H.--P. Cheng, R. N. Barnett, and U. Landman,
   Phys. Rev. B {\bf 48}, 1820 (1993).

\bibitem{Akola}
   J. Akola and M. Manninen,
   Phys. Rev. B {\bf 65}, 245424 (2002).    

\bibitem{Kumar}
   V. Kumar, Phys. Rev. B {\bf 60}, 2916 (1999).  

\bibitem{Aguado-impurity}
   A. Aguado, L. E. Gonz{\'a}lez, and J. L{\'o}pez,
   J. Phys. Chem. B {\bf 108}, 11722 (2004).

\bibitem{Mottet}
   C. Mottet, G. Rossi, F. Baletto, and R. Ferrando,
   Phys. Rev. Lett. {\bf 95}, 035501 (2005).

\bibitem{Si16Ti}
   S. Zorriassatein, K. Joshi, and D. G. Kanhere,
   http://arxiv.org/ps/cond--mat/0609196.    

\bibitem{BOMD}
   M. C. Payne, M. P. Teter, D. C. Allan, T. A. Arias, and J. D. Joannopoulos,
   Rev. Mod. Phys. {\bf 64}, 1045 (1992).

\bibitem{Pseudo}
   D. Vanderbilt,
   Phys. Rev. B {\bf 41}, R7892 (1990).

\bibitem{VASP}
   {\it Vienna Ab initio Simulation Package} (VASP),
   Teachnishe Universit{\" a}t Wien, 1999.

\bibitem{SiSn}
   A. Vichare, D. G. Kanhere, and S. A. Blundell,
   Phys. Rev. B {\bf 64}, 045408 (2001);
   S. Krishnamurty, K. Joshi, D. G. Kanhere, and S. A. Blundell, 
   Phys. Rev. B {\bf 73}, 045419 (2006). 

\bibitem{Li-Fournier}
   R. Fournier, J.--B.--Y.  Cheng, and A. Wong,
   J. Chem. Phys. {\bf 119}, 9444 (2003).

\bibitem{Li10}
   R. O. Jones, A. I. Lichtenstein, and J. Hutter,
   J. Chem. Phys. {\bf 106}, 4566 (1996).

\end{thebibliography}
\end{document}